\documentclass[aps,prl,twocolumn,groupedaddress,amsmath,nofootinbib]{revtex4-1}

\usepackage{amssymb}
\usepackage{amsfonts}
\usepackage{amssymb}
\usepackage{amsmath}
\usepackage[dvips]{graphicx}
\usepackage{color}
\usepackage{amsfonts}
\usepackage[dvips]{graphicx}
\usepackage{epstopdf}
\usepackage{amsmath}
\usepackage{ulem}
\usepackage{color}
\usepackage{amsmath,amssymb,graphics,epsfig,color,times,bbm}
\usepackage{verbatim}
\usepackage{multirow}
\usepackage{upgreek}

\usepackage{txfonts}
\usepackage{hyperref}

\begin{document}

\title{Semi-device-independent quantum random number generator with a
broadband squeezed state of light}
\author{Jialin Cheng$^{1}$}
\thanks{These authors contributed equally to this work.}
\author{Shaocong Liang$^{1}$}
\thanks{These authors contributed equally to this work.}
\author{Jiliang Qin$^{1,2}$}
\author{Jiatong Li$^{1}$}
\author{Zhihui Yan$^{1,2}$}
\author{Xiaojun Jia$^{1,2}$}
\email{jiaxj@sxu.edu.cn}
\author{Changde Xie$^{1,2}$}
\author{Kunchi Peng$^{1,2}$}

\affiliation{$^1$State Key Laboratory of Quantum Optics and Quantum Optics Devices,
Institute of Opto-Electronics, Shanxi University, Taiyuan 030006, China\\
$^2$Collaborative Innovation Center of Extreme Optics, Shanxi University,
Taiyuan, Shanxi 030006, China\\
Email: jiaxj@sxu.edu.cn
}

\begin{abstract}
Random numbers are a basic ingredient
of simulation algorithms and cryptography, and play a significant part in
computer simulation and information processing. One prominent feature of a
squeezed light is its lower fluctuation and more randomness in a pair of orthogonal oriented quadratures, thus it prompts a significant application in
not only quantum information and quantum precision measurement but
also an excellent entropy source for true random number generation. Here we
report a generation of a high-efficiency semi-device-independent quantum
random number based on a broadband squeezed light, where a reliable
randomness source is unnecessary and a noisy local oscillator is allowed for homodyne detection. The equivalent generation of private random bits is at a rate of 580.7 Mbps. In addition, the use of squeezed light at 1.3 $\upmu$m enables the transmission of entropy sources and local oscillators at the metropolitan scale, thus expanding the potential applications of quantum random number generators based on non-classical state of light.
\end{abstract}

\keywords{Quantum information, Nonlinear optics, Quantum random number generator}

\maketitle

\section*{INTRODUCTION}

As an important resource, random number plays an important role in many fields, such as
statistical sampling, computer simulation, cryptography and lotteries. The random numbers based on classical computer algorithms are usually
known as pseudo-random numbers. In order to generate true random numbers, the inherent randomness of quantum effects is
exploited, which is usually called quantum random number generation \cite{7Herrero,8Ma}. Compared with pseudo-random number generators based on computer algorithms or a lack of knowledge of the original state, quantum random number generator (QRNG) exploits a quantum stochastic process as the sources of randomness, which guarantees the generation of truly random numbers. The usual QRNGs include two parts, entropy source and corresponding measure-and-process system. The most simple and effective entropy source is the polarization of arbitrary polarization
state. Series of work about QRNGs have been realized by measuring the polarization of single photon with a single-photon detector \cite{7Herrero,8Ma}. However, the performances of the usual direct-detection system including dead time and detection efficiency seriously limit the rate of random number
generation. Then several ways have been proposed and realized to enhance
the random number generation rate, such as measuring the temporal or spatial
mode of a photon \cite{Yan10}, the distribution of photon number of a laser
pulse \cite{Apple} and so on \cite{Wahl,LouQ}. The quadrature of the quantum state or the phase noise of the laser could be a good source for random number generation.

The usage of real quantum state as an entropy source
can obviously enhance the security of random number generation obviously and generate true random numbers \cite{8Ma}. Thus, there are
multifarious QRNGs based on various quantum resources, in which the most
high-profile is device-independent (DI) QRNG \cite{Pironio,Bierhorst,Liu2,LiuW,LiMH,Acin}. A fully DI QRNG can be achieved solely by
observing a violation of the Bell's inequality \cite{Bell,CHSH}, which does
not rely on any physical implementations and the characterizations of the
realistic instruments \cite{8Ma}. In the DI QRNG protocol, the loophole-free Bell violation implies that the predetermined quantum measurements are performed while the outcomes are fundamentally unpredictable \cite{Pivoluska,WuD}, for example, the typical Bell-type game \cite{Liu2}. This protocol does not assume anything inside the devices used apart from space-like separation and free choice \cite{TBBTC}, such approach can be seen as DI. However, the low generation rate and demanding physical conditions seriously limit the practical application of DI QRNG. Other QRNGs, based on the conventional protocols, are usually achieved in prepare-and-measure scenarios. In these protocols, some devices for preparation or measurement are assumed to be trusted in some condition. For instance, the source-independent (SI) QRNG \cite{13Marangon,14Avesani,ZhengZ,LinX,AvesaniM} and measurement-device-independent (MDI) QRNG \cite{17Cao,258Piran,YuY} need complete characterizations of the measurement device and source respectively, which are commonly called semi-device-independent (SDI) QRNG, while there are also protocols based on other weak assumptions, such as the dimension of the produced quantum states \cite{LiHWY,LunghiT}, their overlap \cite{BraskJB,TebyanianH}, their energy \cite{HimbeeckTV,RuscaD}, and so on. By contrast, SDI QRNG generates faster random bits with less stringent experimental requirements. In this condition, the rate of generated random number is significantly improved by
appropriately assuming and modeling the source or the measurement device, which may meet the requirements for practical applications.

\begin{figure}[htbp]
	\begin{center}
		\includegraphics[width=80mm]{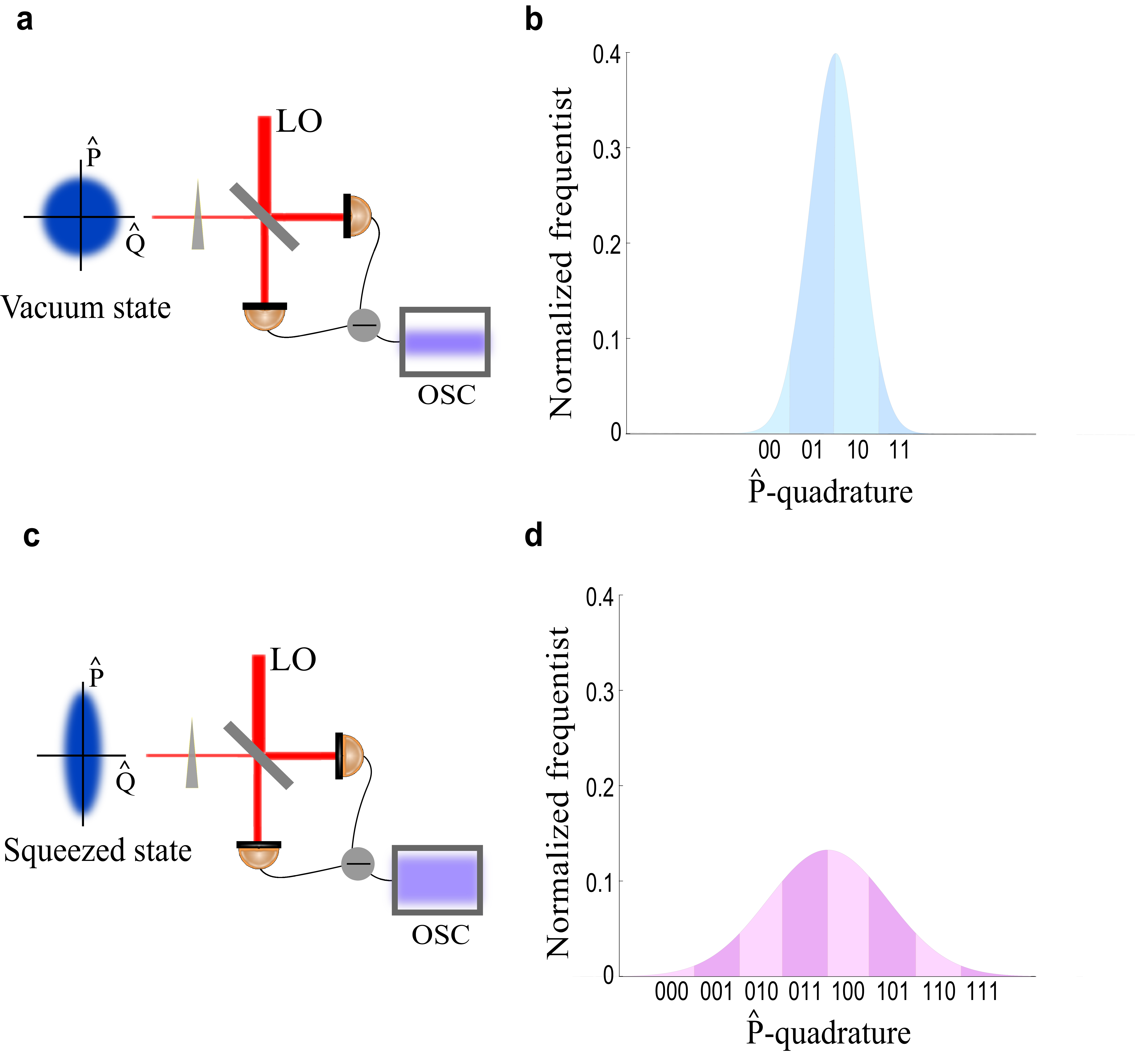}
	\end{center}
	\caption{\label{Fig_1} \textbf{The generation of raw random numbers with different entropy sources in CVs.} \textbf{a-b,} the vacuum state as the entropy source. \textbf{c-d,} A squeezed state as the entropy source. The signal field (the vacuum or squeezed state) and a strong LO interfere on a 50/50 beam splitter. Then the difference signal from two photodetectors is digitized and recorded with an OSC. A flatter and wider noise distribution of the anti-squeezed quadrature of the squeezed state contains more randomness that can be extracted.}
\end{figure}

With the emergence of SDI quantum random number protocols, high-speed QRNGs can now be designed based on quadratures measurement of continuous-variable (CV) quantum states. Because the high-performance macroscopic phtodetector is used for the measurement, the generation rate of QRNG can be greatly improved in this frame \cite{253Weed}. The common and economical way is measuring the quadrature of the vacuum or a squeezed state with a high-efficiency balanced homodyne
detector (BHD) as shown in Fig. ~\ref{Fig_1}. The vacuum state can be easily prepared and
measured and the measurement performance can also be simply compensated by
optimizing the local oscillator (LO) power. By allocating the analog result of a single measurement from the homodyne detector to multiple bits during digitization, the system can generate multiple random bits per measurement, enabling high-speed quantum random number generation with relatively simple equipment. The quadratures can be measured via BHD with a strong LO, and the resulting output current is filtered, digitized and recorded with an oscilloscope (OSC). The finite resolution of the OSC limits the discretization of the continuous variable measurements, thus it is necessary to divide the measurement range of the OSC into $2^{n}$ bins. Considering a trial with a total of $N$ measurements, the raw random numbers are obtained when $N$ outcomes fall into different bins that are coded with $n$ bits ($n$ is determined by the resolution of the OSC). The discretization of quantum measurement corresponds to the partitions ${\{(k\delta o,(k+1)\delta o]\}}_{k=-\infty}^{k=+\infty}$ with integer $k$, where $\delta o$ is the precision of the quadrature $\hat{O}$ ($\hat{O}=\{\hat{P},\hat{Q}\}$) measurement. In fact, this is a process that Alice applies the coarse-grained POVMs ${\{\hat{O}^{k}_{\delta o}}\}$ with elements $\hat{O}^{k}_{\delta o}={{\int_{k\delta o}^{(k+1)\delta o}}}do|o\rangle \langle o|$ and stores the outcomes $o_k$. Compared with the vacuum state, a squeezed state with an orthogonal oriented quadrature noise below the shot noise limit (SNL) can also be used as an entropy source for QRNG \cite{16Michel,Zhu3}. Furthermore, the entropy of the anti-squeezed quadrature of the squeezed state is obviously larger than that of the vacuum state, which means more randomness can be extracted when the noise of the anti-squeezed quadrature is utilized for the QRNG as shown in Fig. ~\ref{Fig_1}c-~\ref{Fig_1}d.

In this article, we experimentally demonstrate a SDI QRNG exploiting a broadband squeezed state of light. Based on a short optical parametric amplifier (OPA), a squeezed state ranging from 3.0 MHz to about 200.0 MHz is obtained and used as the entropy source of the QRNG. Moreover, we introduce a security analysis of the LO (as a part of the measurement device) by parameter measurement of detection system, real-time monitoring of the LO fluctuations to eliminate some eavesdropper's interference and attacks on the measurement device by means of a noisy LO. The usage of the untrusted source and noisy LO at 1.3 $\upmu $m greatly improves the portability of the SDI QRNG. Therefore, its security is higher than usual SI QRNGs and the amount of extractable private randomness per unit bandwidth is significantly improved.

\section*{RESULTS}

\begin{figure*}[htbp]
	\begin{center}
		\includegraphics[width=15cm]{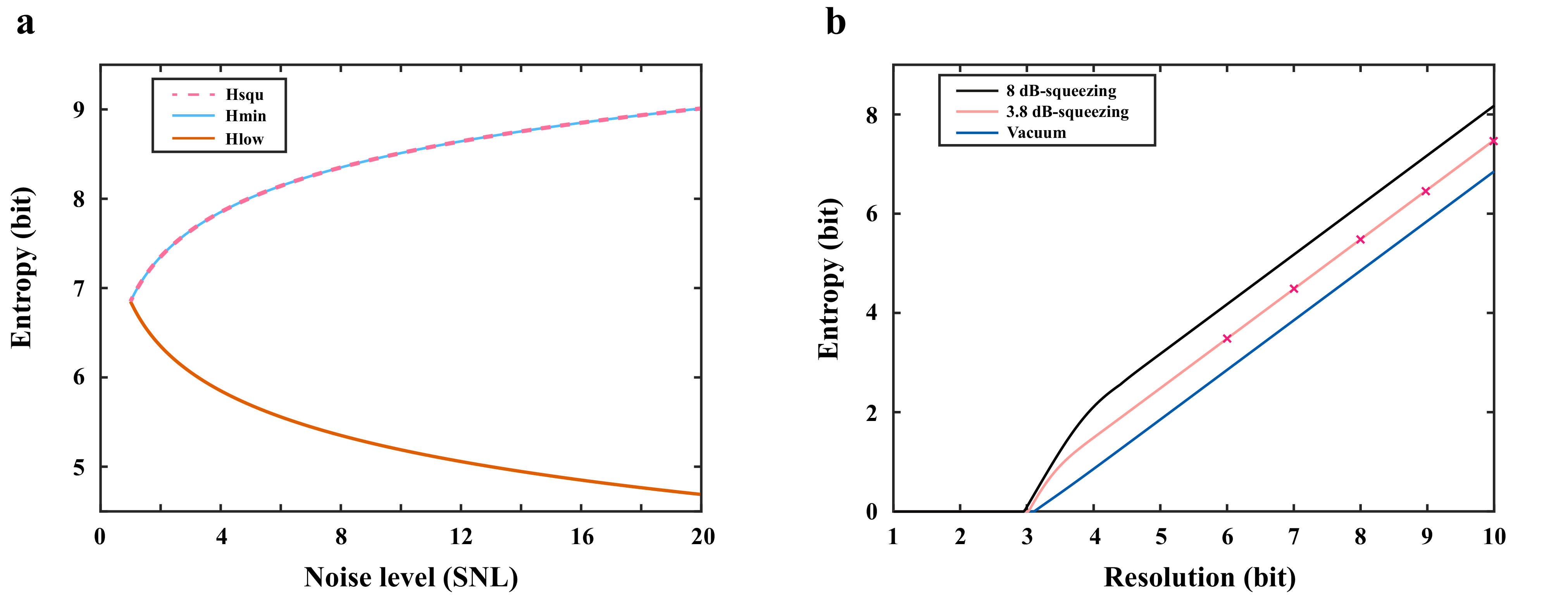}
	\end{center}
	\caption{\label{Fig_2} \textbf{The entropies of different states as functions of the noise level and resolution.} \textbf{a,} The dashed line ($H_{\rm squ}$) represents the lower bound on the conditional min-entropies of the anti-squeezed quadrature of the squeezed state, the light blue ($H_{\rm min}$) and orange curves ($H_{\rm low}$) represent the min-entropy of the thermal state and the lower bound on the conditional min-entropy of the thermal state respectively. All the curves are estimated with a precision $\delta =0.01536$, and the noise levels are set in units of shot noise. The noise level represents the additional noise for thermal state and pure quantum noise for squeezed state, respectively. \textbf{b,} The lower bound on the conditional min-entropy versus the resolution of the oscilloscope for the vacuum state, a 3.8 dB squeezed state and 8.0 dB squeezed state. The precision is set as
		$\delta =0.01536$. The marks show the amount of extractable secure bits in our experiment.}
\end{figure*}

\noindent \textbf{The security of QRNG and entropic uncertainty principle}

To ensure the security of a QRNG, it is paramount to safeguard the generated quantum random numbers against eavesdropping attempts. For instance, when Alice employs a QRNG, she must consider potential eavesdropper Eve, who may employ both classical and quantum methods to eavesdrop the generated random numbers.

The overall security of a QRNG can be defined as follows. Let $\rho_{OE}$ denote the classical-quantum state of Alice's random bits $O$ and a possible eavesdropper $E$. Such a state can always be written as
\begin{equation}
\rho _{OE} = {\textstyle \sum_{o\in O}}p(o)|o\rangle \langle o| \otimes \rho _{E}^{o},
\end{equation}%
where $p(o)$ is the probability distribution of the random bits, $\rho _{E}^{o}$ is the state of the adversary's system given that $O$ takes the value $o$. For $\epsilon >0$, we then call the protocol $\epsilon$-secure if for any eavesdropper $E$
\begin{equation}
\frac{1}{2} \left \| \rho_{OE}- \rho_{U}\otimes \rho_{E}\right \|_{1} \le \epsilon
\end{equation}%
holds. Here, $\left \| \cdot \right \|_{1}$ is the trace norm, $\rho_{U}$ is the uniform distribution over $O$ and $\rho_{E}$ is the reduced state of $\rho_{OE}$.

The entropic uncertainty principle (EUP), which originates from the well-known Heisenberg uncertainty principle, establishes a lower bound on the (quantum) conditional min-entropy of measuring quadrature for a CV quantum state \cite{9Furrer,10Berta,11Vallone,23Furrer}. In fact, randomness determined by general min-entropy does not eliminate the possibility of eavesdropping, which may lead to partial insecure randomness offered by eavesdropper in the bits after post-processing. However, the conditional min-entropy gives the amount of private and secure randomness in the presence of eavesdroppers  \cite{13Marangon,16Michel}. In accordance with the convention of security analysis in QRNG, Alice
denotes the legitimate user generating random numbers, while Eve, the
malicious eavesdropper, tries to guess correctly the outcomes of measurements by adopting an optimal strategy. The EUP plays a central role in SI QRNG protocol for CV quantum systems. The early EUP \cite{ColesPJ}
\begin{eqnarray}
H(P)+H(Q)\ge \log_{2}{\frac{1}{c_{0}}}
\end{eqnarray}%
originates from the famous Heisenberg uncertainty principle $\bigtriangleup P\cdot \bigtriangleup Q\ge \frac{1}{2}\left | \left \langle \left [ P, Q \right ]  \right \rangle  \right |$, where the term $\frac{1}{c_{0}}$ quantifies the complementarity of the observables. In Ref. \cite{9Furrer}, the form of EUP for QRNG described by quantum conditional entropy is given by
\begin{eqnarray}
H_{\min }(P_{\delta p}|E)+ H_{\max}(Q_{\delta q})\geq -\log _{2}c(\delta q,\delta p)
\end{eqnarray}%
with
\begin{equation}
c(\delta q,\delta p)=\frac{1}{2\pi } \delta q\delta pS_{0}^{\left ( 1 \right ) } (1,\frac{\delta q\delta p}{4} )^{2}
\end{equation}%
where $H_{\min }(P_{\delta p}|E)$ is the conditional min-entropy of $P_{\delta p}$. $H_{\max }(Q_{\delta q})$ is the max-entropy expressing Alice's lack of knowledge about outcomes of measuring quadrature $\hat{Q}$. Here $\hat{P}$, $\hat{Q}$ are defined as data quadrature and check quadrature, respectively. The term $%
c(\delta q,\delta p)$ denotes ``incompatibility''\ of the two quadratures $%
\hat{P}$ and $\hat{Q}$. $\delta p$ and $\delta q$ represent the measurement precision of quadratures $\hat{P}$ and $\hat{Q}$ in phase space with an actual digital device, and normally $\delta p = \delta q = \delta$ is set for the continuity and convenience of quadratures measurement. $S_{0}^{(1)}$ is the 0th radial prolate spheroidal wave function of the first kind \cite{9Furrer}.

The smooth min- and max-entropies are usually exploited in operational settings, since they result from the non-smoothed versions by an optimization procedure over states close to the original state \cite{23Furrer}. The generalized residual hash lemma indicates the existence of a family of hash functions such that for any state $\rho _{AE}$, secure random numbers that are almost uniform and independent of $E$ can be obtained. Therefore, exploiting the variation of the conditional min-entropy, smooth min-entropy, the ultimate generation rate of the secure random bits is further bounded as \cite{13Marangon,24Eberle}:
\begin{eqnarray}
H_{\min }^{\epsilon }(P_{\delta p}|E) &\geq &-\log _{2}c(\delta q,\delta
p)-H_{\max}(Q_{\delta q}))-\frac{1}{\sqrt{n_{p}}}\Delta   \notag \\
&\equiv &H_{\rm low}^{\epsilon }(P_{\delta p}|E)
\end{eqnarray}
with
\begin{equation}
\Delta =4\sqrt{\log _{2}(\frac{2}{\epsilon ^{2}})}\log _{2}(2^{1+\frac{%
H_{\max }(Q_{\delta q})}{2}}+1),
\end{equation}%
where $H_{\min }^{\epsilon }(P_{\delta p}|E)$ is the smooth min-entropy, $n_{p}$ is the number of measurements for quadrature $\hat{P}$ and $\epsilon $ is the security parameter. The protocol is called $\epsilon$-secure, which means that it is  $\epsilon$-close to true random number with uniform distribution and independent of $E$.

\noindent \textbf{Advantages of a broadband squeezed light as entropy source}

Following the SI QRNG protocol \cite{13Marangon,16Michel} and ignoring hardware and post-processing, the key factors affecting the generation rate of QRNG mainly include the precision of coarse-grained measurement $\delta $, the bandwidth of measured signal and the purity of quantum states (more details can be found in Supplementary Note 3). Compared with the vacuum, exploiting a broadband squeezed light as the entropy source can enhance the generation rate due to more (secure) randomness from the outcomes of the anti-squeezed quadrature measurements. As shown in Fig. ~\ref{Fig_1}, the outcome of anti-squeezed quadrature is a flatter and wider distribution. Moreover, an extended measurement bandwidth can significantly improve the generation rate of random numbers (see Supplementary Note 3). In short, it is an appreciable improvement by exploiting a wider-bandwidth and higher-level squeezed state as the entropy source.
The min-entropy of the thermal state, (quantum) conditional min-entropy of the thermal state and the conditional min-entropy of the anti-squeezed quadrature of squeezed state as functions of the noise level have been theoretically analyzed respectively, as shown in Fig. ~\ref{Fig_2}a. The noise level represents the additional noise for thermal state and pure quantum noise for squeezed state, respectively. More details about the additional noise can be found in Supplementary Note 3. With the increase of the additional noise in the measured quadrature, the min-entropy of the thermal state $H_{\rm min}$ increases, in which the quadrature seems to contain more randomness. However, the increased randomness cannot be used to generate secure random numbers because the additional noise decreases the purity of the state and may be insecure. The additional noise may be correlated with Eve and give side information to Eve. The content of secure randomness can be estimated by the conditional min-entropy, therefore it is clear that the extractable secure randomness of the thermal state $H_{\rm low}$ decreases with increasing noise level. In contrast, the increasing noise of the anti-squeezed quadrature of the squeezed state, which comes from the property of the quantum state and is pure quantum noise so that it can generate secure random bits. The lower bound on the conditional min-entropy of the anti-squeezed quadrature of the squeezed state is the same as the min-entropy of the thermal state with identical noise level.

Without regard to the sample size, the lower bound on the conditional min-entropy as a function of the resolution of the OSC is shown in Fig. ~\ref{Fig_2}b, where the vacuum, a 3.8 dB-squeezed state (in our case) and an 8.0 dB-squeezed state are chosen as entropy sources, respectively. The precision is set as $\delta =0.01536$ in our experiment. It can be seen that no secure random bits can be extracted when the resolution is very low for all these quantum states. When the resolution increases to a certain value, the lower bound on the conditional min-entropy increases linearly with the resolution.

\begin{figure*}[htbp]
	\begin{center}
		\includegraphics[width=17cm]{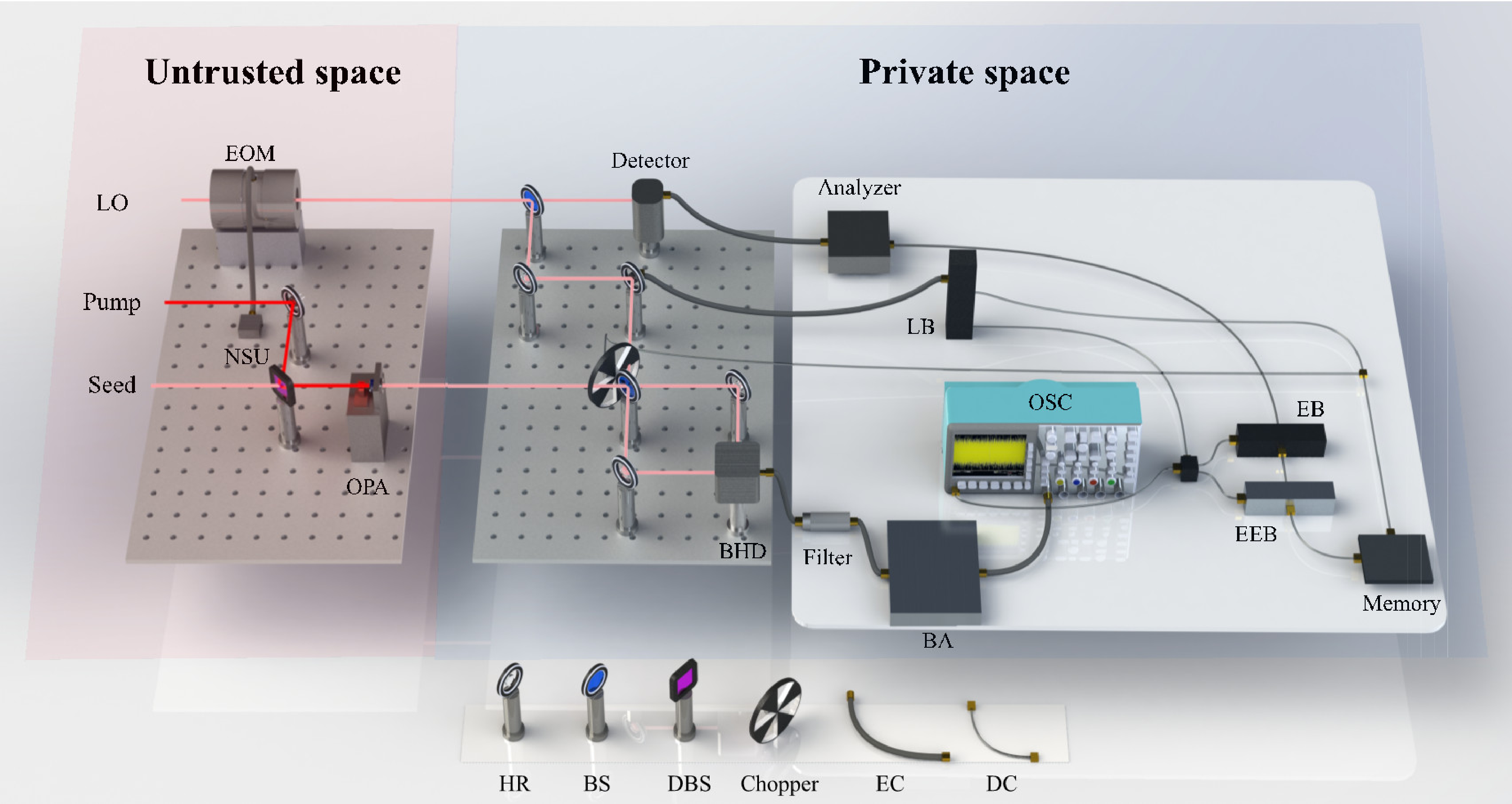}
	\end{center}
	\caption{\label{Fig_3} \textbf{Schematic of experimental configuration for SDI QRNG.} The whole space can be divided into an untrusted space and a private space. A broadband squeezed state is generated form the OPA, and detected with a BHD and the LO transmitted from the untrusted space. The check quadrature, electronic noise and the LO fluctuation are used to calibrate the conditional min-entropy. The data quadrature is used to generate raw random bits. Very few terminal random bits are injected into the locking box and chopper to determine the quadrature type and measure electronic noise respectively. EC: electric cable; DC: data cable; HR: mirror with high reflectivity; BS: 50/50 beam splitter; DBS: dichroic beam splitter; OPA: optical parametric amplifier; BHD: balanced homodyne detector; BA: broadband amplifier; LB: locking box; OSC: oscilloscope; EEB: entropy extraction box; EB: entropy estimation box; NSU: noise source unit; EOM: electro-optic modulator.}
\end{figure*}
\noindent \textbf{The security analysis of LO}

There is not a delicate characterization of the noisy LO in CV SI QRNG framework. Simply setting a power bound on the LO and incessantly monitoring the LO power may not eliminate most security threats to the measurement device. A notable threat is that Eve modulates the LO and adds extra controlled classical noise
as side information to the measured quadrature furtively, which leads to an
overestimation of the content of private randomness. Although there is no such threat for a ideal BHD \cite{27Huang}. The security
of ultimate private bits is also compromised since Alice cannot notice the existence of
Eve by monitoring LO power in this case.

In order to close this backdoor in the measurement device of QRNG, it is necessary to record real-time LO fluctuations and perform a further calibration of the lower bound on the conditional min-entropy in CV SI QRNG scenario, which ensures that this protocol is capable of eliminating some side-channel attacks on the measurement device. On the basis of the theory derived in Supplementary Note 4, we quantify the untrusted noise caused by
LO fluctuation in units of shot
noise. The total noise $\sigma _{\rm M-vac}^{2}$, $\sigma _{\rm M-squ}^{2}$ and $\sigma _{\rm M-ant}^{2}$ measured on quadratures of the vacuum, the squeezed quadrature $\hat{Q}$ and anti-squeezed quadrature $\hat{P}$ of the squeezed state are respectively given by:%
\begin{eqnarray}
\sigma _{\rm M-vac}^{2} &=&\sigma _{\rm SNL}^{2}+\sigma _{\rm LO}^{2}+\sigma _{\rm E}^{2}, \notag \\
\sigma _{\rm M-squ}^{2} &=&\sigma _{\rm squ}^{2}+\sigma _{\rm LO}^{2}+\sigma_{\rm E}^{2},  \notag \\
\sigma _{\rm M-ant}^{2} &=&\sigma _{\rm ant}^{2}+\sigma _{\rm LO}^{2}+\sigma_{\rm E}^{2}.
\end{eqnarray}
The terms $%
\sigma _{\rm SNL}^{2}$, $\sigma _{\rm LO}^{2}$, and $\sigma _{\rm E}^{2}$ denote the measured pure vacuum noise, the untrusted noise introduced by LO
fluctuation and the intrinsic electronic noise of the measurement device, respectively. $\sigma _{\rm squ}^{2}$, $\sigma _{\rm ant}^{2}$ denote the noise of squeezed and anti-squeezed quadratures, respectively.

The electronic noise can be measured by blocking the signal field and LO. Through carefully measuring the relevant parameters of the homodyne detector and recording the LO fluctuations in real time, the untrusted noise introduced by LO fluctuation can be obtained. Then the lower bound on the conditional min-entropy can be further calibrated.

\noindent \textbf{Experimental setup.}

\begin{figure*}[htbp]
	\begin{center}
		\includegraphics[width=15cm]{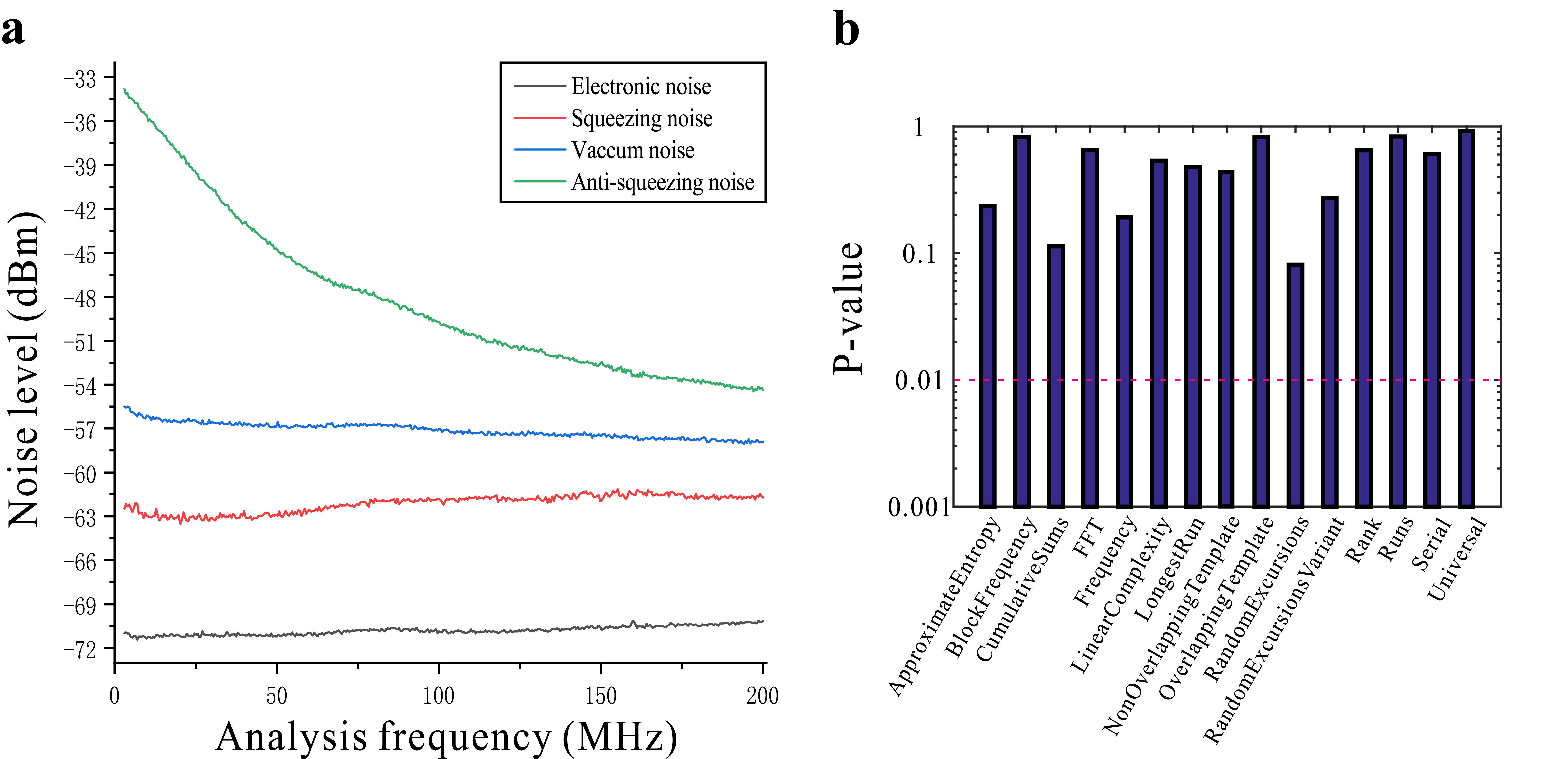}
	\end{center}
	\caption{\label{Fig_4} \textbf{Experimental results.} \textbf{a,} The measured noise power of the broadband squeezed state of light. Black curve: electronic noise level. Blue curve: shot-noise-limit (SNL) with the total LO power of 7.8 mW. Red curve: noise power of squeezed quadrature amplitude $\hat{Q}$. Green curve: noise power of anti-squeezed quadrature phase $\hat{P}$. The measured frequency is from 3.0 MHz to 200.0 MHz, and a $6.5\pm0.2$ dB of squeezing at 3.0 MHz and $3.8\pm0.2$ dB of squeezing at 200.0 MHz below the corresponding SNL is observed. \textbf{b,} The results of a typical run of the NIST test. In the case of multiple tests in a category, the smallest have been reported.}
\end{figure*}

The schematic of experimental setup of SDI QRNG with a broadband squeezed state is shown in Fig. ~\ref{Fig_3}, which consists of two parts, an untrusted space and a private space. The two spaces can be connected by standard telecommunication fiber for the portability of the QRNG. The untrusted space contains an uncertified broadband squeezed state and an incompletely characterized LO. The LO can be noisy since Eve may modulate the LO using an electro-optic modulator (EOM) with a controlled noise source unit (NSU) for side information. The laser at 1342 nm and 671 nm from a dual-wavelength laser (not shown in Fig. ~\ref{Fig_3}) are injected into an OPA to generate a broadband squeezed state of light. Eventually, the measured result for the noise power of broadband squeezed
state of light is shown as Fig. ~\ref{Fig_4}a. A $\hat{Q}$-squeezed state with $6.5 \pm 0.2$ dB squeezing at 3.0 MHz and $3.8\pm0.2$ dB squeezing at 200.0 MHz is prepared. More details about the OPA can be found in the Methods section.

The second part is a trusted data acquisition and processing system in the private space. The LO fluctuations are monitored and analyzed in real time via a detector in order to resist the attacks on the LO. The relative phase between the LO and signal field is locked at $\uppi/2$ and randomly switched to 0 via a lock box (LB) according to the random seed, which determines the obtained data type, the data quadrature $\hat{P}$ for raw random numbers and the check quadrature $\hat{Q}$ for conditional min-entropy estimation respectively. The chopper is used to randomly block the signal field and LO to measure the electronic noise. The signals are filtered, amplified and acquired by filters, a broadband amplifier (BA) and an OSC. The untrusted noise introduced by LO fluctuation, the measured check quadrature and electronic noise are fed into the conditional min-entropy estimation box (EB) to calibrate the entropy extraction box (EEB). The measured data quadrature is fed into the EEB to extract secure random bits. A random seed from the terminal random bits in the memory is injected into the locking box and chopper to determine the relative phase of LO and signal field and measure electronic noise respectively.

Compared with the previous SI QRNG protocol \cite{16Michel}, a security
analysis of the noisy LO is used for eliminating some
Eve's interference and attacks on the detectors in our protocol, which acts as an
essential process for improving overall security. In the beginning
of the experiment, the various noise levels of the LO with
different powers over the whole analysis frequency range are measured and
the relevant parameters of the devices inside private space are
calibrated, and then the precision $\delta$ is estimated. The max-entropy $H_{\max}\left( Q_{\delta q}\right)$ is calculated by the recorded data of check quadrature using
frequentist estimator, while the data of the other quadrature are used to generate random numbers. This protocol adopts an extremely conservative approach to security, treating all classical noise and unnecessary quantum noise as impurities within the input state see Supplementary Note 3. Any observed ``mixing''\ is attributed to a specific quantum eavesdropping strategy to safeguard the security of QRNG. According to real-time feedback of the $H_{\max }\left(Q_{\delta q}\right) $, the electronic noise of the detector and the untrusted noise caused by the LO fluctuation, the lower bound on the quantum conditional min-entropy can be estimated. The
Toeplitz-matrix hashing algorithm is applied to extract the private random
bits from the raw random numbers. Abort protocol
whenever the data are out of the range of the devices or the
homodyne detector is saturated. Finally, the generation rate of secure
random bits of our protocol is 580.7 Mbps.

It is noteworthy that we need to emphasize a trivial assumption that these internal devices and hardware in the private space of Fig. ~\ref{Fig_3} are absolutely secure, and there is no
possibility that Eve changes the reliable parts of the
experimental setup (the reflectivity of the beam splitter, the quantum efficiency of
photodiodes and the circuitry part of the detector, etc.) and even
surreptitiously installs backdoor malware. However, Eve knows all of these
parameters and the running process of the setup and protocol, and is capable of regulating the temperature, the electric and magnetic fields around the private space
to control and manipulate the electronic noise for side information during the generation of random numbers.

\section*{DISCUSSION}

Unlike most CV quantum key distribution (QKD) schemes \cite{FhX}, we use continuous LO instead of optical pulses, and there is no need for synchronization of RF signals for source preparation and quadrature measurement, which greatly simplify the experimental system and avoid many attacks exist in CV-QKD. In addition, the LO fluctuation is inappreciable in our SDI QRNG, whose effect on the measured noise is small to distinguish with our
parameters of the homodyne detector as calculated in the Methods section. However, if the ratio of the reflectivity to transmission is far from 1:1 for a substandard BHD, very little classical fluctuation of the LO will give much side information to Eve. Alice has no way to combat
such an eavesdropping strategy by monitoring LO power and
aborting the protocol with noticed detector saturation. Moreover, we have taken note of the antenna attack problem which has been recently proposed as an area of concern \cite{SmithPR}. It may be beneficial to eliminate this antenna attack by monitoring the detector for saturation and random measurements of the detector's electronics noise.

SI QRNG shows a good prospect in both classical and quantum
communication networks. The securities and generation rates of SI QRNGs
need more detailed description and further improvement. Our SDI QRNG
protocol aims to improve not only security but generation
rate of QRNG. A broadband squeezed light is exploited to improve generation rate with a maintained measured bandwidth. Based on the traditional SI protocol, a security analysis of the LO is introduced for increasing the overall security. Finally, the generation rate of secure random bits is estimated to 580.7 Mbps, which can meet the requirements of an ordinary short-range QKD system and a quantum secure direct communication system \cite{Hujy,ZhangHET}. In addition, the use of squeezed light at 1.3 $\upmu $m enables the transmission of entropy sources and LOs at the metropolitan scale, and makes it possible for the generation of the random numbers certified by Bell's theorem. Recently, microring resonators are exploited as optical parametric oscillator \cite{LuJ,BWA,OkY} and some quantum light sources have been produced on a photonic chip \cite{Zhaoy,Cas,Singh,luoWET,LiXET}, which mean a smaller SDI QRNG and wider squeezing bandwidth are achievable. Meanwhile, there have been some reports of random number generations with a photonic integrated chip \cite{BaiBB,AbellanC} and an optical chip for self-testing quantum random number generation with a silicon photomultiplier \cite{LeoneN}. On-chip quantum random number generation is a major leap towards miniaturization and integration of QRNG. Moreover, a broadband squeezed light has inestimable advantages in information processing and pure state preparation,
and shows extraordinary potential in quantum information \cite{30Kashiwazaki,31Politi,32Masada} and quantum measurement \cite{33Yu} due to
its wide bandwidth as well as higher purity at higher frequency.

\section*{METHODS}
\noindent \textbf{Details about the experimental setup}

The OPA contains a $1\times 2\times 6$ $\rm mm^{3}$ periodically poled KTiOPO$_4$ (PPKTP) crystal and a piezo-actuated concave mirror with a radius of curvature of 150 mm which works as the output coupler. The length of the cavity is about 10 mm. The front surface of the PPKTP is coated with a film of reflectivity $R>99.9\%$ at 1342 nm and transmission $T=85.3\%$ at 671 nm. The rear surface of the PPKTP is coated with anti-reflection (AR) at both 1342 nm and
671 nm. The output coupler is coated with reflectivity $R=88.2\%$ at 1342 nm and
$R=99.9\%$ at 671 nm. The finesse of the OPA is about 25. Compared with our previous squeezed light sources \cite{28Hou,YanZJia,Zuoxj}, the extended bandwidth of the OPA is achieved mainly by
redesigning the structure of the OPA and shortening the cavity length. With a film-coated PPKTP crystal, the usual input mirror can be omitted in this semi-monolithic configuration \cite{5ZhouYY}. Then the half width at half maximum of the OPA can be estimated to about 202 MHz because the FSR of the cavity is approximately 10.1 GHz. When the length of the cavity and working condition are well controlled with corresponding locking systems, the intracavity losses are $0.3\%$ and the parametric gain is about 20 at the pump power of 96 mW. Eventually a bright squeezed state with the bandwidth of over 200.0 MHz is obtained.

The homodyne detection is achieved with a homemade detector
included two photodiodes FD150 (Fermionics Opto-Technology company) to meet the requirements of wide measured bandwidth and high
quantum efficiency for broadband squeezed states. The interference signal of the signal field and the LO is exploited as the error signal to achieve the locking of the quadrature phase (i.e., the data quadrature), and provides a high-speed switching signal to the fiber-coupled electro-optic modulator (not shown in Fig. ~\ref{Fig_3}) for achieving quadrature amplitude switching according to the random seed. The direct current of
the detector is the interference signal, while the alternating current is used to acquire data for further steps. Furthermore, all frequencies of the modulating signals for frequency stabilization and phase locking are set
below 1 MHz to offer an appropriate frequency range for the SDI QRNG
protocol. For instance, the frequency stabilization of the mode cleaners is achieved by exploiting a homemade locking box with a modulating signal of 10 kHz. The propagation efficiency, fringe visibility, and photodiodes quantum efficiency are 0.99, 0.99 and 0.88, respectively. The electronic noise level is about -13 dB compared to the SNL with a LO intensity of 7.8 mW at 200.0 MHz. The
results of measured noise power of the generated broadband squeezed state of light in Fig. ~\ref{Fig_4}a are
obtained after filtering out low-frequency noise via a 3.0 MHz high-pass filter. There is substantial classical noise including relaxation oscillation noise and modulating signals existing in the range of 0 to 3 MHz, thus the measured results in this range have to be discarded for the generation of secure random bits. On the other hand, only squeezing of 3.8 dB in the full frequency range (from 3.0 MHz to 200.0 MHz) is used to ensure the security of QRNG in the most conservative situation. Otherwise, the amount of secure randomness will be overestimated.

\noindent \textbf{Experimental results and data processing}

The nominal resolution of the OSC is over 11 bits. The signals are acquired at a rate of 1.5 GSamples per second and then downsampled to 200 MSamples per second. The relative phase between the LO and signal beam is randomly switched to quadrature $\hat{Q}$ and collect the check data in every 20 $\mu $s, and the electronic noise is randomly measured in every 0.4 s. We set a ratio of 1/20 between the total measurements and the check measurements, and so does the electronic noise.

In a round of random number generation, the measured vacuum noise contains (5.75 $\pm $ 0.02)\% of electronic
noise and less than $10^{-4}$ \% of the untrusted noise introduced by the LO fluctuation. The precision in phase space is estimated to be $\delta =\left( 1.536\pm 0.002\right) \times 10^{-2}$, a detailed process can be found in Supplementary Note 5. The duration of randomly measuring check quadrature and electronic noise are set as 1$\upmu$s and $0.02$ s respectively. $2\times 10^{5}$ samples are obtained and the security parameter is set as $\epsilon
=10^{-6}$. The lower smooth conditional min-entropy is estimated to be $H_{\min }^{\epsilon }(P_{\delta p}|E)=7.21\pm 0.04$ bits per
sample. Taking into account the effective number of bits of the OSC,
we use a conservative bit depth of 6 bits, and the entropy is reduced to $%
3.32\pm 0.04$ bits. A $2$ Mb $\times$ $656$ kb Toeplitz matrix hashing is
applied to extract random bits. In addition, we need to consume very few private bits for the data-type switching. Finally, the equivalent generation of private random bits is at a rate greater than 580.7 Mbps.

For an undisturbed LO, it can be regarded as a coherent state. However, if Eve
modulates the LO, the LO fluctuations will be superimposed
on the measured noise with a ratio of $K_{1}/K_{2}$ (see Supplementary Note 4 for more details). For instance, if the reflectivity and transmission of the substandard beam splitter are about $21.2\%$ and $78.9\%$ respectively, and the LO noise modulated by Eve increases by 3 dB for an eavesdropping strategy, then the measured vacuum noise is not pure and contains $50\%$ classical noise, i.e., $\sigma _{\rm LO}^{2}=\sigma _{\rm SNL}^{2}$. Eventually, one can estimate that $31.02\%$ of the extracted random bits are not secure and private.

In order to check for the randomness in terminal private bits, we test
them with the NIST suite \cite{44Bass}. The results of a typical run are reported in Fig. ~\ref{Fig_4}b. In the case of multiple tests in a
category, the smallest has been reported. The private bits pass all the NIST statistical tests.

In addition, the squeezing of CV quantum state rapidly decays with losses, and the generation rate decreases with transmission distance. However, as long as the squeezed state can be maintained in transmitting process, the bit rate of a squeezed state is better than that of the vacuum. With the single mode loss of 0.35 dB per kilometer at 1.3 $\upmu $m in fiber, the squeezing level reduces to 1.4 dB at 200 MHz when the transmission distance is about 9.27 km. Therefore, there is still obvious advantage over the vacuum in generation rates.

\hspace*{\fill}

\textbf{DATA AVAILABILITY}

These data for experimental results can be found at https://github.com/chengjialin007/bssQRNG. The code is available from the corresponding author on request.

\textbf{ACKNOWLEDGMENTS}

This work was supported by the National Natural
Science Foundation of China (Grants No. 61925503, No. 62122044, No.
12147215, and No. 11834010), the Key Project of the National Key
R\&D program of China (Grant No. 2016YFA0301402), the Program for the Innovative Talents
of Higher Education Institutions of Shanxi, the Program for the Outstanding Innovative Teams
of Higher Learning Institutions of Shanxi and the fund for Shanxi \textquotedblleft 1331
Project\textquotedblright\ Key Subjects Construction.

\textbf{COMPETING INTERESTS}

The authors declare no competing interests.

\textbf{AUTHOR CONTRIBUTIONS}

X.J. and C.X. conceived the original idea. J.C., S.L., Z.Y., X.J., and K.P. designed the experiment. J.C., S.L., J.Q., and J.L. constructed and performed the experiment. J.C., S.L., and X.J. accomplished theoretical calculation and the data analysis. J.C., Z.Y., X.J., C.X., and K.P. wrote the paper. All the authors reviewed the manuscript. J.C. and S.L. contributed equally to this work.

\newpage

\end{document}